\documentclass{PoS}

\usepackage{graphicx}
\usepackage{cite}
\usepackage{bm}
\usepackage{graphicx}
\usepackage{pgf}
\usepackage{amsmath}
\usepackage{xcolor}

\usepackage{tikz}
\usetikzlibrary{shapes,arrows}
\usepackage{picture}

\tikzstyle{decision} = [diamond, draw, 
    text width=4.5em, text badly centered, node distance=3cm, inner sep=0pt]
\tikzstyle{block} = [rectangle, draw, 
    text width=8em, text centered, rounded corners, minimum height=4em]
\tikzstyle{line} = [draw, -latex']
\tikzstyle{cloud} = [draw, ellipse, node distance=3cm, text width=6em,
    minimum height=2em]

\def\ba{\begin{eqnarray}}
\def\ea{\end{eqnarray}}
\def\la{\langle}
\def\ra{\rangle}
\def\bs{\bm} 
\def\wt{\widetilde}

\def\bt{\bs{T}}
\def\e{\epsilon}
\def\nn{\nonumber}
\def\NF{N_F}
\def\b#1{\bar{#1}}
\def\b#1{\bar{#1}}

\def\hb#1{\hat{\bar{#1}}}

\def\dsigma{{\rm d} \hat\sigma}

\title{Colourful antenna subtraction for gluon scattering}

\ShortTitle{Colourful antenna subtraction for gluon scattering}

\author{{James Currie}
        \\
        E-mail: \email{jcurrie@physik.uzh.ch}}

\abstract{In this talk I discuss the application and generalization of the antenna subtraction method to processes involving incoherent interferences of partial amplitudes, which are generically present for the sub-leading colour contributions to processes involving more than five partons. The approach makes use of the known infrared (IR) singularity structure of one- and two-loop matrix elements to guide the construction of the subtraction terms. A set of integrated dipoles are defined which can be used to express the poles of one- and two-loop matrix elements in terms of integrated antennae.  The unintegrated counterparts of these subtraction terms are then inferred to construct the double real and real-virtual subtraction terms.  The method has been tested by computing the NNLO sub-leading colour contribution dijet production via gluon scattering. The double real and real virtual matrix elements for this process can be written purely in terms of incoherent interferences and so the successful removal of all singularities and divergences demonstrates the ability of the antenna subtraction method to handle general sub-leading colour contributions.}

\FullConference{11th International Symposium on Radiative Corrections (Applications of Quantum Field Theory to Phenomenology) (RADCOR 2013),\\
		22-27 September 2013\\
		Lumley Castle Hotel, Durham, UK }

\begin{document}

\section{Introduction}

With the successful operation of the LHC in recent years, the exploration of the terascale is moving apace.  The discovery of the Higgs boson by the ATLAS and CMS collaborations~\cite{Aad:2012tfa,Chatrchyan:2012ufa} has set the keystone of the Standard Model into place, whereas new physics beyond the Standard Model (BSM) is yet to reveal itself.  As BSM physics retreats to less readily accessible regions of parameter space it is natural to focus our attention on high precision Standard Model phenomenology. High precision phenomenology is essential for realistically simulating the hadron collider environment, especially when working within the framework of perturbation theory where many physical features of hadron collisions only enter the game through higher order perturbative corrections. 

The excellent performance of the LHC, in particular with respect to the jet energy scale (JES) calibration~\cite{Schouten:2012en,Kirschenmann:2012toa}, demands next-to-next-to leading order (NNLO) perturbative corrections for a variety of Standard Model processes and so in recent years some considerable effort of the theory community has been spent developing methods to perform NNLO calculations. Some of these methods are restricted to colourless final-states or fully inclusive calculations~\cite{babishiggs,Catani:2007vq}. Others have pursued a fully numerical approach following the innovation of the \texttt{STRIPPER} method~\cite{Czakon:2010td} which has subsequently been applied to various studies of top pair~\cite{czakontop4} and Higgs plus one jet production~\cite{Boughezal:2013uia}. The antenna subtraction method was developed for electron-positron collisions and successfully applied to the process $e^{+}e^{-}\to3j$ at NNLO~\cite{our3j1}.  In recent years this method has been extended to accommodate hadronic initial states~\cite{Daleo:2006xa,Daleo:2009yj,Boughezal:2010mc,GehrmannDeRidder:2012ja,Gehrmann:2011wi} and successfully applied to dijet production for a set of partonic channels~\cite{GehrmannDeRidder:2013mf,Currie:2013vh,Currie:2013dwa}. 

Antenna subtraction is currently the only available method for performing NNLO calculations containing generic coloured initial and final states with analytic pole cancellation. For this reason it is important to develop the method in full generality so that it may deal with a broad range of processes.  One aspect of antenna subtraction which has not received widespread attention is the issue of sub-leading colour corrections to observables. In this context, the phrase \emph{sub-leading colour} refers generically to any terms suppressed by at least $1/N^{2}$ relative to the leading term when expanding the squared matrix element as a series in $N$. The essential issue for these corrections is that the matrix element cannot, in general, be written as a sum of squared coherent partial amplitudes, where the colour ordered amplitude and its conjugate share the same colour ordering, e.g.,
\ba
M_{n}(\sigma(1),\cdots,\sigma(n))&=&{\cal{M}}_{n}^{\dagger}(\sigma(1),\cdots,\sigma(n)){\cal{M}}_{n}(\sigma(1),\cdots,\sigma(n)).
\ea
At leading colour, due to colour coherence, the matrix element is guaranteed to be in this form.  Similarly, for processes with a small number of external legs ($n\leq5$) it can often be arranged such that the matrix element is formed from squared coherent partial amplitudes. In these cases, antenna subtraction is particularly well suited to remove the IR divergences from the matrix element. The ability of antennae to remove such divergences can be traced back to the fact that antennae are defined to be squared partial amplitudes. The sub-leading colour contributions for a sufficiently complicated process, such as dijet production, therefore pose a question;  Can the antenna subtraction method, as it currently stands, be successfully applied to the kind of incoherent interferences generically found at sub-leading colour?

To address this question we consider the sub-leading colour corrections to dijet production via the ``gluons only'' approximation, in which all external states are gluons and no quarks propagate in any loops, i.e. $\NF=0$. This process is convenient for studying the behaviour of incoherent interferences because the double real and real-virtual corrections can be written entirely in terms of such interferences.

The IR factorization behaviour of incoherent interferences is more complicated than that of squared coherent partial amplitudes. One approach to constructing the subtraction terms is to examine the IR behaviour of the matrix element in each limit and match those limits to the known IR behaviour of the antenna functions. For a general process this can be cumbersome as there are many single and double unresolved limits to consider, each of which must be considered separately. 

An alternative approach is to use the IR pole structure of the one- and two-loop matrix elements, clearly exposed by Catani~\cite{Catani:1998bh}, to guide the construction of the subtraction term.  In order to do this a set of \emph{integrated dipoles}~\cite{Currie:2013vh} are introduced, constructed from integrated antenna functions and mass factorization kernels, such that the singularities of these dipoles reproduce the virtual IR structure elucidated by Catani's formalism. Synthesising the colour stripped integrated dipoles with the colour space formalism allows single and double virtual subtraction terms to be constructed for general processes. Furthermore, there exists a clear and unambiguous link between these integrated dipoles and their unintegrated counterparts.  This link suggests that the colour space formalism can also be used to guide the construction of the unintegrated subtraction terms. In a recent paper~\cite{Currie:2013dwa} both approaches were taken and found to be complementary; in this talk I will briefly outline the colourful antenna subtraction approach used for the calculation of the sub-leading colour contribution to jet production via gluon scattering.

\section{Integrated antenna dipoles}

The IR pole structure is particularly transparent when formulated in terms of insertion operators acting on states in colour space~\cite{Catani:1998bh}. The principle observation of colourful antenna subtraction is that an analogous set of single ($\ell=1$) and double ($\ell=2$) unresolved insertion operators can be defined in terms of the known integrated antennae, and where necessary, mass factorization kernels,
\ba
\bm{J}^{(\ell)}(\e)&=&2\sum_{(i,j)}\bm{J}_{2}^{(\ell)}(i,j)~\bt_{i}\cdot\bt_{j},
\ea
where the sum runs over all pairs of partons. The type of integrated dipole is determined by the parton species and whether the partons in the dipole are in the initial- or final-state. The poles of the $\bs{J}^{(2)}(\e)$ operator are not in a one-to-one correspondence with the relevant piece of Catani's $\bs{I}^{(2)}(\e)$ operator due to the fact that the antenna dipoles are inherently real functions and also there are finite differences between the $\bs{J}^{(1)}(\e)$ and $\bs{I}^{(1)}(\e)$ operators; however, these differences drop out in the combination of operators used to construct the double virtual subtraction term.

In the ``gluons only'' approximation the colour stripped single unresolved dipoles are given by~\cite{Currie:2013dwa},
\ba
\bs{J}_{2}^{(1)}(1_{g},2_{g})&=&\frac{1}{3}{\cal{F}}_{3}^{0}(s_{{1}{2}}),\label{eq:j21def1}\\
\bs{J}_{2}^{(1)}(\hb{1}_{g},2_{g})&=&\frac{1}{2}{\cal{F}}_{3,g}^{0}(s_{\b{1}{2}})-\frac{1}{2}{\Gamma}_{gg}^{(1)}(x_{1})\delta(1-x_{2}),\label{eq:j21def2}\\
\bs{J}_{2}^{(1)}(\hb{1}_{g},\hb{2}_{g})&=&{\cal{F}}_{3,gg}^{0}(s_{\b{1}\b{2}})-\frac{1}{2}{\Gamma}_{gg}^{(1)}(x_{1})\delta(1-x_{2})-\frac{1}{2}{\Gamma}_{gg}^{(1)}(x_{2})\delta(1-x_{1})\label{eq:j21def3},
\ea
where the hatted arguments denote initial-state gluons. Similarly, the colour stripped double unresolved integrated dipoles are given by~\cite{Currie:2013dwa},
\ba
\bs{J}_{2}^{(2)}(1_{g},2_{g})&=&\frac{1}{4}{\cal{F}}_{4}^{0}(s_{{1}{2}})+\frac{1}{3}{\cal{F}}_{3}^{1}(s_{{1}{2}})+\frac{1}{3}\frac{b_{0}}{\epsilon}{\cal{F}}_{3}^{0}(s_{{1}{2}})\bigg(\biggl(\frac{|s_{{1}{2}}|}{\mu^{2}}\biggr)^{-\epsilon}-1\bigg)\nn\\
&-&\frac{1}{9}\big[{\cal{F}}_{3}^{0}\otimes{\cal{F}}_{3}^{0}\big](s_{12}),\\
\bs{J}_{2}^{(2)}(\hb{1}_{g},2_{g})&=&\frac{1}{2}{\cal{F}}_{4,g}^{0}(s_{\b{1}{2}})+\frac{1}{2}{\cal{F}}_{3,g}^{1}(s_{\b{1}{2}})+\frac{1}{2}\frac{b_{0}}{\epsilon}{\cal{F}}_{3,g}^{0}(s_{\b{1}{2}})\bigg(\biggl(\frac{|s_{\b{1}{2}}|}{\mu^{2}}\biggr)^{-\epsilon}-1\bigg)\nn\\
&-&\frac{1}{4}\big[{\cal{F}}_{3,g}^{0}\otimes{\cal{F}}_{3,g}^{0}\big](s_{\b{1}2})-\frac{1}{2}\overline{\Gamma}_{gg}^{(2)}(x_{1})\delta(1-x_{2}),\\
\bs{J}_{2}^{(2)}(\hb{1}_{g},\hb{2}_{g})&=&{\cal{F}}_{4,gg}^{0,\rm{adj}}(s_{\b{1}\b{2}})+\frac{1}{2}{\cal{F}}_{4,gg}^{0,\rm{n.adj}}(s_{\b{1}\b{2}})+{\cal{F}}_{3,gg}^{1}(s_{\b{1}\b{2}})+\frac{b_{0}}{\epsilon}{\cal{F}}_{3,gg}^{0}(s_{\b{1}\b{2}})\bigg(\biggl(\frac{|s_{\b{1}\b{2}}|}{\mu^{2}}\biggr)^{-\epsilon}-1\bigg)\nn\\
&-&\big[{\cal{F}}_{3,gg}^{0}\otimes{\cal{F}}_{3,gg}^{0}\big](s_{\b{1}\b{2}})-\frac{1}{2}\overline{\Gamma}_{gg}^{(2)}(x_{1})\delta(1-x_{2})-\frac{1}{2}\overline{\Gamma}_{gg}^{(2)}(x_{2})\delta(1-x_{1}),
\ea
where the integrated antennae and mass factorization kernels may be found in~\cite{GehrmannDeRidder:2012dg}. 

\section{Subtraction terms}

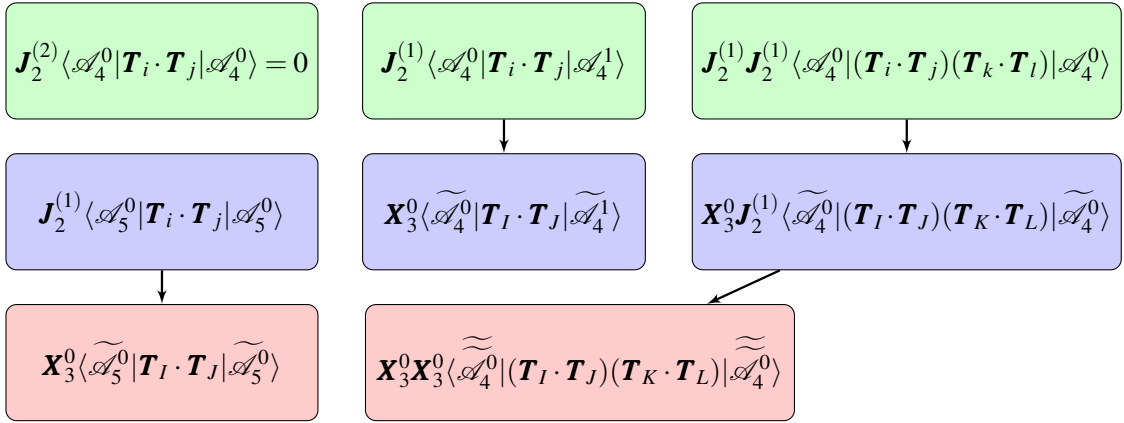
\begin{figure}[h]
\begin{tikzpicture}[node distance = 2cm, auto]
    \node [block, text width=10em, fill=green!20] (vv1) {$\bs{J}_{2}^{(2)}\la{\cal A}_{4}^{0}|\bt_{i}\cdot\bt_{j}|{\cal A}_{4}^{0}\ra=0$};
    \node [block, text width=9em, fill=green!20, right of=vv1, node distance=4.5cm] (vv2) {$\bs{J}_{2}^{(1)}\la{\cal A}_{4}^{0}|\bt_{i}\cdot\bt_{j}|{\cal A}_{4}^{1}\ra$};
    \node [block, text width=14em, fill=green!20, right of=vv2, node distance=5.3cm] (vv3) {$\bs{J}_{2}^{(1)}\bs{J}_{2}^{(1)}\la{\cal{A}}_{4}^{0}|(\bt_{i}\cdot\bt_{j})(\bt_{k}\cdot\bt_{l})|{\cal{A}}_{4}^{0}\ra$};
    \node [block, text width=10em, below of=vv1, fill=blue!20, node distance=2cm] (rv1) {$\bs{J}_{2}^{(1)}\la{\cal A}_{5}^{0}|\bt_{i}\cdot\bt_{j}|{\cal A}_{5}^{0}\ra$};
    \node [block, text width=9em, below of=vv2, fill=blue!20, node distance=2cm] (rv2) {$\bs{X}_{3}^{0}\la\wt{{\cal A}_{4}^{0}}|\bt_{I}\cdot\bt_{J}|\wt{{\cal A}_{4}^{1}}\ra$};
    \node [block, text width=14em, fill=blue!20, below of=vv3, node distance=2cm] (rv3) {$\bs{X}_{3}^{0}\bs{J}_{2}^{(1)}\la\wt{{\cal{A}}_{4}^{0}}|(\bt_{I}\cdot\bt_{J})(\bt_{K}\cdot\bt_{L})|\wt{{\cal{A}}_{4}^{0}}\ra$};
    \node [block, text width=10em, below of=rv1, fill=red!20, node distance=2cm] (rr1) {$\bs{X}_{3}^{0}\la\wt{{\cal A}_{5}^{0}}|\bt_{I}\cdot\bt_{J}|\wt{{\cal A}_{5}^{0}}\ra$};
    \node [block, text width=14em, fill=red!20, right of=rr1, node distance=5.5cm] (rr2) {$\bs{X}_{3}^{0}\bs{X}_{3}^{0}\la\wt{\wt{{\cal{A}}_{4}^{0}}}|(\bt_{I}\cdot\bt_{J})(\bt_{K}\cdot\bt_{L})|\wt{\wt{{\cal{A}}_{4}^{0}}}\ra$};
    \path [line, line width=0.3mm] (vv2) -- (rv2);
    \path [line, line width=0.3mm] (vv3) -- (rv3);
    \path [line, line width=0.3mm] (rv1) -- (rr1);
    \path [line, line width=0.3mm] (rv3) -- (rr2);
\end{tikzpicture}
\caption{Schematic depiction of the links between subtraction terms at the double virtual (green), real-virtual (blue) and double real (red) levels for gluon scattering at sub-leading colour. Arrows denote links between terms related by analytic integration such that the unintegrated subtraction terms may be inferred from their integrated counterparts.}
\label{fig:flow}
\end{figure}

To successfully compute the NNLO correction to jet production, we must construct the double real, real-virtual and double virtual subtraction terms, denoted by $\dsigma_{NNLO}^{S}$, $\dsigma_{NNLO}^{T}$ and $\dsigma_{NNLO}^{U}$ respectively.  Many of these subtraction terms are connected to each other via analytic integration; the integration of a piece of one generates a piece of another. For reference throughout this section, a schematic depiction of the links between terms is shown in Fig.~\ref{fig:flow}.

In colourful antenna subtraction the natural place to start is with the double virtual subtraction term. This subtraction term must remove all poles in $\e=2-d/2$ from the two-loop matrix element so that the four-dimensional remainder can be numerically integrated. The IR pole structure of the two-loop matrix element can be correctly reproduced using the single and double unresolved insertion operators and thus removed using the colourful double virtual subtraction term,
\ba
\lefteqn{\dsigma_{NNLO}^{U}=-{\cal N}_{LO}\left(\frac{\alpha_s}{2\pi}\right)^2\bar{C}(\epsilon)^2
\int \frac{{\rm d}z_1}{z_1}\frac{{\rm d}z_2}{z_2}~{\rm d}\Phi_{2}(p_3,p_4;z_{1}{p}_1,z_{2}{p}_2)~{\cal{J}}_{2}^{(2)}(p_{3},p_{4})}\nn\\
&\Big(&2{\rm{Re}}\la{\cal A}_{4}^{0}|\bs{J}^{(1)}(\e)|{\cal A}_{4}^{1}\ra-\la{\cal{A}}_{4}^{0}|[\bs{J}^{(1)}(\e)\otimes\bs{J}^{(1)}(\e)]|{\cal A}_{4}^{0}\ra+N\la{\cal A}_{4}^{0}|\bs{J}^{(2)}(\e)|{\cal A}_{4}^{0}\ra\Big),\label{eq:vvpoles}
\ea
where $|{\cal A}_{4}^{\ell}\ra$ is the $\ell$-loop four gluon amplitude as an abstract vector in colour space, ${\cal{J}}_{2}^{(2)}$ is the jet function, $\bar{C}(\epsilon)=8\pi^2 C(\e)=(4\pi)^{\e}e^{-\e\gamma}$ and ${\cal{N}}_{LO}$ takes into account any overall factors arising from initial-state colour and spin averaging which are not explicitly included in the evaluation of the colour charge sandwiches. It is useful to define the leading colour, $\mathbb{LC}$, and sub-leading colour, $\mathbb{SLC}$, projectors which single out the leading and sub-leading colour contributions.  Evaluating the colour algebra reveals that,
\ba
\mathbb{SLC}\big(N\la{\cal A}_{4}^{0}|\bt_{i}\cdot\bt_{j}|{\cal A}_{4}^{0}\ra\big)&=&0,
\ea
and so the third term in Eq.~(\ref{eq:vvpoles}), involving $\bs{J}^{(2)}(\e)$, does not contribute to the sub-leading colour subtraction term. This has implications for the double real and real-virtual subtraction terms as it implies there are no four-parton antennae in the double real, nor one-loop antennae in the real-virtual subtraction terms. Evaluating the other terms explicitly yields a subtraction term composed of integrated three-parton antennae and reduced matrix elements.\footnote{It is noted that in this subtraction term, as with all formulae for the process, the mass factorization kernels in the integrated dipoles mutually cancel, leaving only integrated antennae.  This is necessary as the mass factorization kernels are only present at leading colour for this process.}

The real-virtual subtraction term contains essentially three components corresponding to the left, centre and right blue boxes in Fig.~\ref{fig:flow}:
\begin{itemize}
\item $\dsigma_{NNLO}^{T,a}$ which removes all explicit poles from the one-loop matrix element.
\item $\dsigma_{NNLO}^{T,b_1}$ which removes all single unresolved divergences from the one-loop matrix element.
\item $\dsigma_{NNLO}^{T,b_{2}}$ and $\dsigma_{NNLO}^{T,c}$ which remove any spurious poles or divergences introduced by the previous two contributions.
\end{itemize}
The poles of the one-loop matrix element are reproduced by the single unresolved insertion operator and so the appropriate subtraction term is given by,
\ba
\dsigma_{NNLO}^{T,a}&=&-{\cal N}_{LO}\left(\frac{\alpha_s}{2\pi}\right)^2\frac{\bar{C}(\epsilon)^2}{C(\e)}
\int \frac{{\rm d}x_1}{x_1}\frac{{\rm d}x_2}{x_2}~{\rm d}\Phi_{3}(p_3,p_4,p_5;x_{1}{p}_1,x_{2}{p}_2)\nn\\
&\times&\mathbb{SLC}\Big(2{\rm{Re}}\la{\cal A}_{5}^{0}|\bs{J}^{(1)}(\e)|{\cal A}_{5}^{0}\ra\Big)~{\cal{J}}_{2}^{(3)}(p_{3},p_{4},p_{5}),\label{eq:rvpoles}
\ea
where the sum over pairs of partons inside the definition of the insertion operator now runs over the set of five external gluons. As shown in Fig.~\ref{fig:flow}, this term is the integrated version of the single unresolved double real subtraction term, $\dsigma_{NNLO}^{S,a}$.

In single unresolved limits, the one-loop matrix element typically factorizes into two contributions: one containing a tree-level singular function multiplied by a one-loop reduced matrix element, and another containing a one-loop singular function multiplied by a tree-level reduced matrix element. 

The one-loop singular functions are approximated by one-loop antennae and so upon integration this set of subtraction terms contributes to the $\bs{J}_{2}^{(2)}$ dipoles. It has already been shown that, for this process, any terms contributing to this dipole are zero at sub-leading colour and so the one-loop antenna component of $\dsigma_{NNLO}^{T,b_{1}}$ is absent.

The only contribution to the double virtual subtraction term that is proportional to the reduced one-loop matrix element is the first term in Eq.~(\ref{eq:vvpoles}). It is from this double virtual subtraction term that $\dsigma_{NNLO}^{T,b_{1}}$ can be inferred.  By replacing the integrated dipole by the relevant unintegrated antenna, summing over all potentially unresolved gluons, and applying the relevant phase space map on the reduced matrix element (fixed by the antenna), we obtain the appropriate real-virtual subtraction term,
\ba
\dsigma_{NNLO}^{T,b_1}&=&-{\cal N}_{LO}\left(\frac{\alpha_s}{2\pi}\right)^2\frac{\bar{C}(\epsilon)^2}{C(\e)}
\int \frac{{\rm d}x_1}{x_1}\frac{{\rm d}x_2}{x_2}~{\rm d}\Phi_{3}(p_3,p_4,p_5;x_{1}{p}_1,x_{2}{p}_2)\nn\\
&\times&2{\rm{Re}}\sum_{i,j,k}X_{3}^{0}(i,j,k)~\mathbb{SLC}\Big(\la\wt{{\cal A}_{4}^{0}}|\bt_{I}\cdot\bt_{K}|\wt{{\cal A}_{4}^{1}}\ra\Big)~{\cal{J}}_{2}^{(2)}(p_{I},p_{K}),\label{eq:dstb1}
\ea
where the type of antenna depends on which partons are involved, e.g. $X_{3}^{0}(\hb{1},j,\hb{2})=F_{3}^{0}(\hb{1},j,\hb{2})$, $X_{3}^{0}(\hb{1},j,k)=f_{3}^{0}(\hb{1},j,k)$, $X_{3}^{0}(i,j,k)=f_{3}^{0}(i,j,k)$.  The antenna is associated with a phase space mapping which maps the three partons in the antenna down to two composite partons $(i,j,k)\to(I,K)$ from which the reduced matrix elements are constructed. The remaining real-virtual subtraction terms, $\dsigma_{NNLO}^{T,b_{2}}$ and $\dsigma_{NNLO}^{T,c}$ are also fixed, either directly or indirectly, by the double virtual subtraction term and so can be constructed in a similar fashion to $\dsigma_{NNLO}^{T,b_1}$.  This process involves introducing a set of subtraction terms which are linked to the double real subtraction term by analytic integration, as shown in Fig.~\ref{fig:flow}.

The double real subtraction term contains a contribution proportional to a five parton reduced matrix element, $\dsigma_{NNLO}^{S,a}$ and terms proportional to four parton reduced matrix elements, $\dsigma_{NNLO}^{S,b,c,d,e}$. The former can be inferred from the only real-virtual subtraction term proportional to a five parton matrix element, $\dsigma_{NNLO}^{T,a}$, by replacing the integrated dipole with the relevant antenna and inducing a phase space map on the reduced matrix element,
\ba
\dsigma_{NNLO}^{S,a}&=&-{\cal N}_{LO}\left(\frac{\alpha_s}{2\pi}\right)^2\frac{\bar{C}(\epsilon)^2}{C(\e)^{2}}
\int{\rm d}\Phi_{4}(p_3,p_4,p_5,p_{6};{p}_1,{p}_2)\nn\\
&\times&2{\rm{Re}}\sum_{i,j,k}X_{3}^{0}(i,j,k)~\mathbb{SLC}\Big(\la\wt{{\cal A}_{5}^{0}}|\bt_{I}\cdot\bt_{K}|\wt{{\cal A}_{5}^{0}}\ra\Big)~{\cal{J}}_{2}^{(3)}(p_{I},p_{K},p_{l}).\label{eq:dstb1}
\ea
The remaining double real subtraction terms are fixed by the remaining terms in the real-virtual and double virtual subtraction terms.  Explicit colour summed formulae for all subtraction terms can be found in~\cite{Currie:2013dwa}.

\section{Numerical results}

By evaluating the subtraction terms expicitly, the poles of the subtraction terms can be analytically cancelled against those of the one- and two-loop matrix elements.  The remaining four dimensional matrix elements and subtraction terms can then be implemented in a parton-level Monte Carlo generator and integrated numerically. 

To test the reliability of the subtraction terms derived for this process we consider the single jet inclusive cross section for proton-proton collisions at a centre of mass energy $\sqrt{s}=8$ TeV.  We apply the anti-$k_{t}$ jet finding algorithm with $R=0.7$ and require jets with $p_{T}\ge80$ GeV at central rapidity $|y|\leq4.4$. The MSTW08NNLO gluon parton distribution functions are used and the renormalization and factorization scales are set to the $p_{T}$ of the leading jet.

\begin{figure}[t]
\centering
\includegraphics[width=0.71\textwidth]{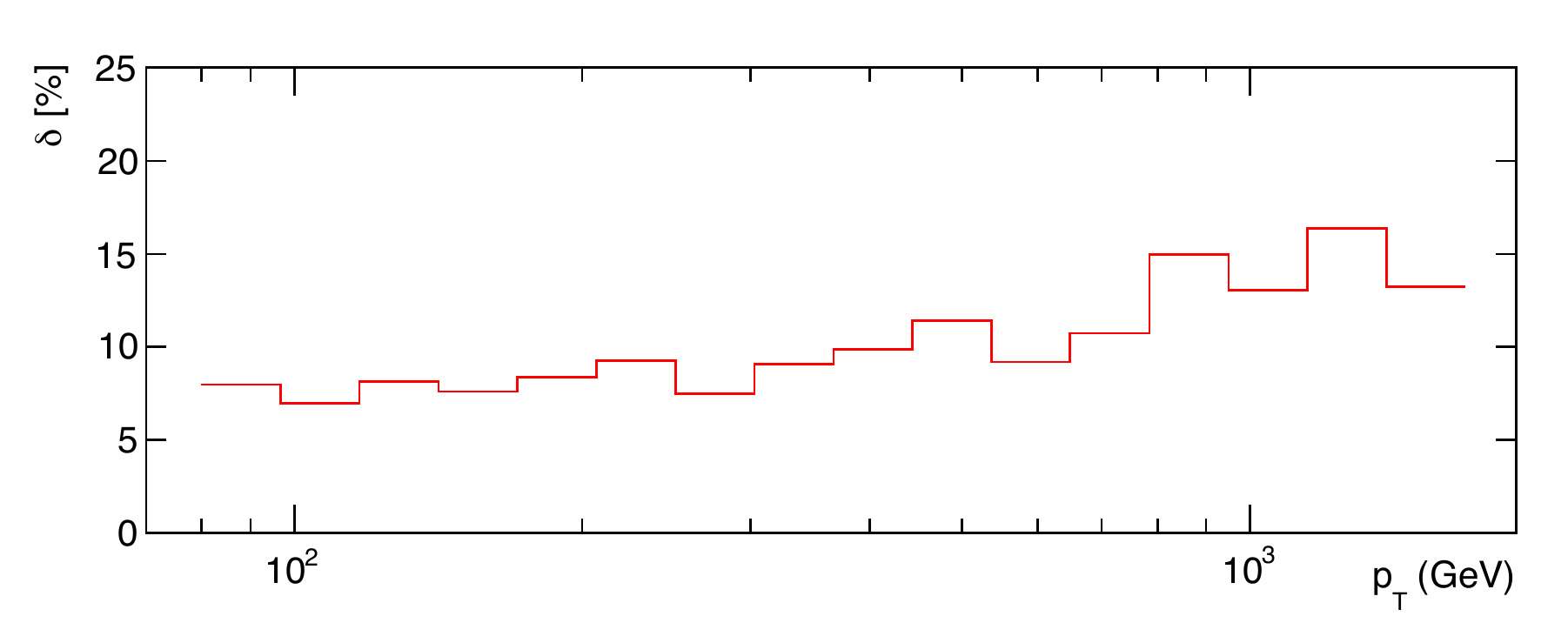}\\\vspace{-0.2cm}
\caption{The percentage contribution of the sub-leading colour to full colour NNLO correction, $\delta$, for the single jet inclusive transverse energy distribution as a function of $p_{T}$.}
  \label{fig:subratio}
\end{figure}

To quantify the size of the sub-leading colour correction to gluon scattering for dijet production we can consider separating the cross section into LO, NLO and NNLO coefficients and further partitioning the NNLO coefficient into leading colour and sub-leading colour contributions,
\ba
{\rm{d}}\sigma&=&\alpha_{s}^{2}A+\alpha_{s}^{3}B+\alpha_{s}^{4}(C_{LC}+C_{SLC}).
\ea
In Fig.~\ref{fig:subratio} we plot the ratio of the sub-leading colour contribution to the full NNLO coefficient as a function of $p_{T}$,
\ba
\delta&=&\frac{C_{SLC}}{C_{LC}+C_{SLC}}.
\ea
Updated full colour single jet inclusive, double differential and dijet invariant mass distributions in the gluons only approximation can be found in~\cite{Currie:2013dwa}.

\section{Summary}

In this talk I have approached the issue of extending antenna subtraction to processes involving incoherent interferences of partial amplitudes, which are generically encountered in sub-leading colour calculations. A set of integrated dipoles can be defined whose pole structure can be matched to the well understood IR pole structure of two-loop matrix elements.  Using these functions, the double virtual subtraction term can be immediately written down. The various components of the double real and real-virtual subtraction terms can also be inferred either from the pole structure of the matrix elements or from the algorithmically generated double virtual subtraction terms.. 

By applying this method to sub-leading colour gluon scattering, it was found that the colourful antenna subtraction terms remove all poles in $\e$ from the double real and real-virtual matrix elements and subtract all IR divergence associated with unresolved partons.  The resulting antenna subtracted cross sections were obtained by numerical integration and produce a positive correction to the leading colour cross section of between 8\% at low $p_{T}$ to 15\% at high $p_{T}$. The magnitude of this result is in line with expectations given that the sub-leading colour correction is suppressed by a factor of $1/N^2$ relative to the leading colour contribution.

The colourful antenna approach treats leading colour and sub-leading colour calculations on an equal footing.  It requires no new antennae, all process-dependent information is encoded through the colour algebra and reduced matrix elements and it has been shown to produce numerically convergent results for a specific calculation. It is therefore a useful extension of the antenna subtraction method, in particular for sub-leading colour calculations and we anticipate its use for additional sub-leading colour corrections to dijet production in the future.

\acknowledgments{I gratefully acknowledge the many useful discussions and collaboration with Jo\~ao Pires, Nigel Glover, Aude Gehrmann-De Ridder and Thomas Gehrmann. This research was supported by the European Commission through the `LHCPhenoNet' Initial Training Network PITN-GA-2010-264564.}

\bibliography{ref}

\end{document}